\documentclass[apjl]{emulateapj}

\shorttitle{Morphology of $z\sim2$ early--types in WFC3/GOODS}
\shortauthors{Cassata et al.}

\begin{document}

\title{The Morphology of Passively Evolving Galaxies at $z\sim 2$ from
{\it HST}/WFC3 Deep Imaging in the Hubble Ultradeep
Field\altaffilmark{1}}

\altaffiltext{1}{Based on observations made with the NASA/ESA \textit{Hubble Space Telescope} operated by the Association of Universities for Research in Astronomy (AURA), Inc., under contract NAS5-26555. }

\author{P. Cassata\altaffilmark{2}, 
M. Giavalisco\altaffilmark{2}, 
Yicheng Guo\altaffilmark{2},
H. Ferguson\altaffilmark{3},
A. Koekemoer\altaffilmark{3},
A. Renzini\altaffilmark{4},
A. Fontana\altaffilmark{5},
S. Salimbeni\altaffilmark{2},
M. Dickinson\altaffilmark{6},
S. Casertano\altaffilmark{3},
C. J. Conselice\altaffilmark{7},
N. Grogin\altaffilmark{3},
J. M. Lotz\altaffilmark{6},
C. Papovich\altaffilmark{8},
R. A. Lucas\altaffilmark{3},
A. Straughn\altaffilmark{9},
Jonathan. P. Gardner\altaffilmark{9},
L. Moustakas\altaffilmark{10}
}

\altaffiltext{2}{Department of Astronomy, University of Massachusetts, Amherst, MA 01003; paolo@astro.umass.edu}
\altaffiltext{3}{Space Telescope Science Institute, 3700 San Martin Drive, Baltimore, MD, 21218}
\altaffiltext{4}{Osservatorio Astronomico di Padova (INAF-OAPD), Vicolo dell'Osservatorio 5, I-35122, Padova, Italy}
\altaffiltext{5}{INAF - Osservatorio Astronomico di Roma, via Frascati 33, Monteporzio-Catone (Roma), I-00040, Italy}
\altaffiltext{6}{NOAO-Tucson, 950 North Cherry Avenue, Tucson, AZ 85719}
\altaffiltext{7}{University of Nottingham, School of Physics and Astronomy, Nottingham NG7 2RD}
\altaffiltext{8}{George P. and Cynthia Woods Mitchell Institute for Fundamental Physics and Astronomy, Department of Physics, Texas A\&M University, College Station, TX 77843-4242, USA}
\altaffiltext{9}{Astrophysics Science Division, Observational Cosmology Laboratory, Goddard Space Flight Center, Code 665, Greenbelt, MD 20771}
\altaffiltext{10}{Jet Propulsion Laboratory, California Institute of Technology, MS 169-327, Pasadena, CA 91109}

\begin{abstract}
We present near--IR images, obtained with the {\it Hubble Space
Telescope} ({\it HST}) and the WFC3/IR camera, of six passive and
massive galaxies at redshift $1.3<z<2.4$ (SSFR$<10^{-2}$ Gyr$^{-1}$;
stellar mass $M\sim10^{11}$ M$_{\odot}$), selected from the Great
Observatories Origins Deep Survey (GOODS). These images, which have a
spatial resolution of $\sim1.5$ kpc, provide the deepest view of the
optical rest--frame morphology of such systems to date. We find that
the light profile of these galaxies is regular and well described by a
S\'ersic model with index typical of today's spheroids. Their size,
however, is generally much smaller than today's early types of similar
stellar mass, with four out of six galaxies having $r_e\sim1$ kpc or
less, in quantitative agreement with previous similar measures made at
rest--frame UV wavelengths. The images reach limiting surface
brightness $\mu\sim$26.5 mag arcsec$^{-2}$ in the F160W bandpass; yet,
there is no evidence of a faint halo in the galaxies of our sample,
even in their stacked image.  We also find that these galaxies have
very weak ``morphological {\cal k}--correction'' between the
rest--frame UV (from the ACS $z$--band), and the rest--frame optical
(WFC3 $H$--band): the S\'ersic index, physical size and overall
morphology are independent or only mildly dependent on the wavelength,
within the errors.

\end{abstract}


\section{Introduction}

Most of the stars observed in today's early--type galaxies have formed
at high redshifts ($z>2$, see Renzini 2006), but the observations have
not yet constrained how that stellar mass has assembled in the
presently observed systems.

Several groups have reported that at least in the range
$M>M\sim10^{11}$ M$_{\odot}$, the stellar mass function of these
high--redshift ``elliptical galaxies'', appears to be rapidly
increasing from $z\sim3$ to $z\sim1$, pointing to this epoch as one
of major assembly for modern bright galaxies (e.g. Bundy~et~al.~2006;
Fontana~et~al.~2006; Arnouts~et~al.~2007; Scarlata~et~al.~2007;
Ilbert~et~al.~2009). A number of mechanisms have been proposed as
drivers of this evolution, from merging to gradual in--situ accretion
(Naab et al. 2009; van Dokkum et al. 2009; Hopkins et al. 2010).

Elliptical galaxies at $z>1.5$ are often observed to be smaller, by factors
$\sim3-5\times$, than their local counterparts of similar stellar mass, and
thus to have $\approx 30-100\times$ higher stellar density (Daddi~et~al.~2005;
Trujillo~et~al.~2006; Trujillo~et~al.~2007; Toft~et~al.~2007;
Zirm~et~al.~2007; Longhetti~et~al.~2007; Cimatti~et~al.~2008,
Van~Dokkum~et.~al.~2008, Buitrago~et~al.~2008, Toft~et~al.~2009), a fact that
still lacks an explanation in terms of an evolutionary mechanism.

Some have suggested that current observations at high redshift could have
missed low surface-brightness halos surrounding the galaxies, due to the
relatively limited sensitivity combined with the $(1+z)^4$ cosmological dimming
(Hopkins~et~al.~2009b; Mancini~et~al.~2009). Others have proposed that surveys
of the local universe (e.g. the SDSS, Stoughton~et~al.~2002) could have missed
a relatively large fraction of very compact, yet massive early--type galaxies
due to the limited ground--based angular resolution (1'' corresponds to a
physical scale of $\sim1$ kpc at $z=0.05$). While recent works
(Valentinuzzi~et~al.~2009; Trujillo~et~al.~2009) reported the identification
of dense and massive galaxies in the local universe, their spatial abundance
appears much smaller than the $z\sim2$ galaxies.

Furthermore, while small high--redshift samples have been imaged with {\it
  HST} and NICMOS at rest--frame optical wavelengths (Buitrago~et~al.~2008;
Trujillo~et~al.~2007; Toft~et~al.~2007; Zirm~et~al.~2007;
Longhetti~et~al.~2007; Van~Dokkum~et.~al.~2008), most of the galaxies have
only been observed in the rest--frame UV with ACS, and the dependence of the
morphology of these galaxies on the wavelength has not been characterized
(Cimatti~et~al.~2008; Daddi~et~al.~2005).

In this letter we use the unprecedented sensitivity and angular
resolution of WFC3/IR images recently acquired in the HUDF to study
the optical rest--frame morphology of a small but well defined sample
of low SSFR galaxies at z$\sim$2. AB magnitudes are used throughout
this paper, and, when needed, a $\Lambda$-CDM world model with
$H_0$=70 km~s$^{-1}$~Mpc$^{-1}$, $\Omega_M=0.3$ and
$\Omega_{\Lambda}=0.7$.

\section{Data Reduction, Sample Selection and Morphological Analysis}
The primary imaging data consist of the first epoch of WFC3/IR
observations acquired by Illingworth~et~al. (2009, program ID=GO11563)
in the HUDF (Beckwith~et~al.~2006), and thus the field is fully
embedded in the GOODS--South field (Giavalisco~et~al.~2004). The
images have been acquired in the F105 (Y), F125W (J) and F160W (H)
filters and covers an area roughly equal to the footprint of the
WFC3/IR camera (2.1 arcmin$^2$). We have carried out our independent
reduction of the raw data and after rejecting images affected by
persistence in the $J$--band, our final stacks reach $1-\sigma$
surface brightness fluctuations of 27.2, 26.6 and 26.3 AB/arcsec$^2$
in the three bands, respectively. We have drizzled the WFC3 images to
0.06'' pixel scale and brought them into registration with the HUDF
ACS and GOODS ones. The size of the PSF of the stacks is 0.18'' in the
H-band. We have then created a multi-wavelength source catalog
(GUTFIT, Grand Unified TFIT Catalog) using the TFIT procedures by
Laidler~et~al.~(2007), which contains PSF--matched photometry in all
the bands acquired by the GOODS multi--facility program, from the U to
the 8$\mu$m IRAC one.

We have selected a sample of high redshift, massive and passive
galaxies using both the BzK color selection by Daddi~et~al.~(2004), as
well as multi--band SED fitting to spectral population synthesis
models. While the BzK technique provides a well characterized
criterion to select relatively massive galaxies with low specific
star--formation rate at $1.4<z<2.5$ , the method naturally has finite
completeness and contamination.  Thus, given the overall limited size
of the sample, we have augmented it with galaxies at $z>1.3$ selected
for being massive and passive according the criteria M$>10^{10}
M/M_{\odot}$ and SSFR$<10^{-2}$ Gyr$^{-1}$ based on fitting their
observed broad--band SEDs to the Charlot \& Bruzual (2007) and
Maraston~(2005) models. We have used Salpeter IMF and lower and upper
mass limits ${\rm 0.1}$ and ${\rm 100}$ M$_{\odot}$, respectively,
obtaining essentially identical results with both libraries. We have
adopted the Calzetti~et~al.~(2000) obscuration law to account for the
possible (modest) presence of dust. Finally, all our sample galaxies
have no detection at 24$\mu$m down to a $1-\sigma$ limit of 5$\mu$Jy,
consistent with predictions for passively evolving galaxies at
$z\sim2$ (Fontana~et~al. 2009).

\begin{figure*}
\begin{center}
\includegraphics[scale=.48]{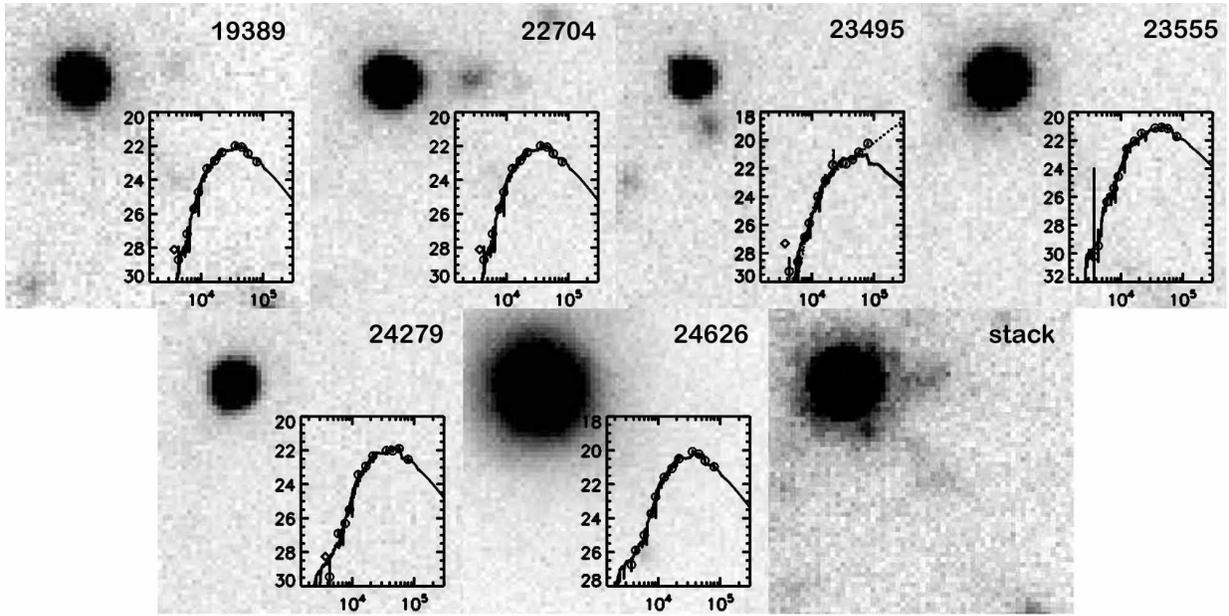}
\caption{The H--band images of the sample galaxies, and the stacked image of
  all but galaxy \# 23495. The size of each panel is $\sim3.5\times3.5
  arcsec^2$. The insets show the observed (data points) and best--fit (curve)
  SED for each galaxy. For object \# 23495 we also show the AGN SED best--fit
  as a dashed curve (see text).}\label{fig1}
\end{center}
\end{figure*}

The final sample consists of six galaxies, four of which have
spectroscopic redshifts, namely \# 22704 and \# 23555 from
Cimatti~et~al.~(2008), \# 24279 from Daddi~et~al.~(2005) and \# 24626
from Vanzella~et~al.~(2008), and two for which we derived accurate
($\Delta z/(1+z)\sim0.05$) photometric redshifts from the GUTFIT
multi-band photometry. Four out of six galaxies satisfy the pBzK
criterion.  Table1~\ref{table:1} lists the sample, including the
parameters of the stellar populations from the fitted procedure, while
Figure ~\ref{fig1} shows the observed (points) and best--fit (curves)
SED together with the H--band images of the galaxies.

\begin{figure*}
\begin{center}
\includegraphics[scale=.8]{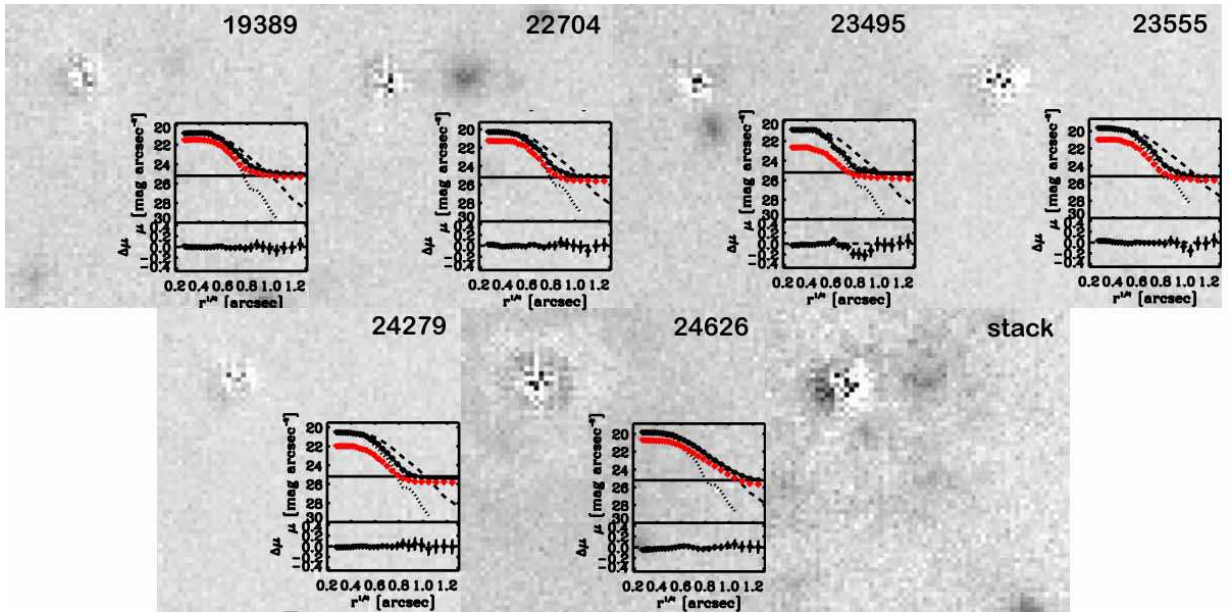}
\caption{The H--band residual images of the sample galaxies and of the
  stacked image obtained subtracting the best--fit S\'ersic model
  (after PSF convolution) from the original images. The insets show
  the observed profiles (diamonds), the best--fit profiles (continuous
  line), the psf profile (dotted line), a S\'ersic model with $r_e=3
  kpc$ and $n_{Sersic}=4$ (dashed line) and the residual profile
  (bottom panel). The sky level is shown as an horizontal line. Black
  symbols refer to the H--band light profile, red diamonds to the
  Y--band one. The red Y-H colors of our galaxies cause the Y-band
  profile to be always below the H-band one.}\label{fig2}
\end{center}
\end{figure*}

We have used the GALFIT package (Peng~et~al.~2002) to fit the light
profile of the galaxies in the $z$--, $Y$--, $J$-- and $H$-- bands to
the S\'ersic model
\begin{equation}
I(r)=I_eexp\left\{-b_n\left[\left(\frac{r}{r_e}\right)^{1/n}-1\right]\right\},
\end{equation}
where $I(r)$ is the surface brightness measured at distance $r$, $I_e$ is the
surface brightness at the effective radius $r_e$ and $b_n$ is a parameter
related to the S\'ersic index $n$. For $n$=1 and $n$=4 the S\'ersic profile
reduces respectively to an exponential or deVaucouleurs profile. Bulge
dominated objects typically have high $n$ values (e.g. $n>2$) and disk
dominated objects have $n$ around unity. Cassata~et~al.~(2005),
Ravindranath~et~al.~(2006) and Cimatti~et~al.~(2008) showed that GALFIT 
yields unbiased estimates of the S\'ersic index and effective radius.

We have obtained the PSF in each passband for use with GALFIT by
averaging six well-exposed, unsaturated stars in the UDF/WFC3
field. For all galaxies, we ran multiple fits with the sky either set
to a fixed, pre-established value or left as a free parameter. We also
experimented with the size of the fitting region around each galaxy,
finding peak--to--peak variations of the S\'ersic indices and $r_e$ at
the 20\% level at the most. The fits in Table~\ref{table:1} use a
$6\times6$ arcsec$^2$ fitting region and free sky.

\section{The morphological properties of galaxies at $z\sim2$.}

\begin{table*}
\caption{The sample of passive galaxies} 
\label{table:1}      
\centering                          
\begin{tabular}{c c c c c c c c c c c c c}        
\hline\hline 
ID & RA & DEC & z$^{(1)}$ &log($M/M_{\odot}$) & log(SSFR) & E(B-V) & Age & $\tau$& S\'ersic & r$_e$ \\ 
& & & & & [Gyr$^{-1}$]& & & Gyr & $H-$band & [kpc] \\ 
\hline 
19389 & 53.135730 & -27.784932 & 1.307p & 10.41 & -5.55 & 0.1 & 3 & 0.2 & 2.99 $^{(2)}$ & 1.02 $^{(2)}$ \\ 
22704 & 53.153799 & -27.774587 & 1.384s & 10.70 & -5.55 & 0.15 & 3 & 0.2 & 2.65 &0.50 \\ 
23495 & 53.158455 & -27.773982 & 2.349p & 11.14 & -3.39 & 0.15 & 2 & 0.2 & $^{(3)}$ & $<0.38$ \\ 
23555 & 53.158810 & -27.797155 & 1.921s & 10.82 & -2.98 & 0.0 & 2 & 0.2 & 1.97 & 0.44 \\ 
24279 & 53.163005 & -27.797655 & 1.980s & 10.63 & -3.39 & 0.15 & 3 & 0.5 & 1.63 & 0.37 \\ 
24626 & 53.165159 & -27.785869 & 1.317s & 11.10 & -2.11 & 0.1 & 3 & 0.5 & 7.42 & 3.69 \\
\hline 
\end{tabular}
\tablecomments{(1): $s$ and $p$ indicate spectroscopic and photometric
  redshift, respectively; (2) modeled with the S\'ersic profile plus the psf
  one; (3) modeled with the psf profile, so the size reported here is just an
  upper limit. } 
\end{table*}

Figure~\ref{fig2} shows the residual images obtained by subtracting
the GALFIT best--fit models (after convolution with the PSF) from the
corresponding images, together with the observed and best-fit model
light profiles (see Table~\ref{table:1}). All galaxies, except \#
24646, are very compact and symmetric, in close resemblance to
present--day spheroidal galaxies. Galaxy \# 24626, the most extended
of our sample, has isophotes that deviate from the elliptical shape,
and the position angle of the isophotes varies as a function of the
distance along the semi--major axis. These irregularities can be the
result of ongoing or recent interactions (see Kormendy~et~al.~2009 and
references therein).

Of the six galaxies, four have S\'ersic index $\gtrsim$2 (\# 19389, \#
22704, \# 23555 and \# 24626). In the case of object \# 19389 we had
to add a central unresolved (PSF) component to the S\'ersic model to
obtain a reasonable fit. This central stellar object contains less
than 10\% of the total light in the object, and may be indicative of
an AGN. This galaxy, however, is not detected in the Chandra Deep
Field South 2--Megasecond Catalog (Luo~et~al.~2008) nor in the radio
VLA maps by Kellermann~et~al.~(2008) and Miller~et~al.~(2008).
Figure~\ref{fig2} also shows that the inner part of the light profile
of galaxy \# 23495 (r$\lesssim1''$) is barely resolved in all bands,
suggesting another AGN. This appears confirmed by the fact that the
galaxy is also detected in the Chandra image, and it has X--ray
luminosity 3.8$\pm0.24\times10^{43} erg/s$ and 5.6$\pm0.6\times10^{43}
erg/s$ in the soft and hard band, respectively; it is not, however,
detected in the VLA maps. With the exception of the IRAC $8~\mu$m data
point, whose value suggests the presence of a hot dust component, the
broad--band SED of this source is otherwise very similar to that of
the other galaxies, and fits made under the assumption that its
luminosity is powered by stars yield ``stellar populations
parameters'' that are essentially the same as the other galaxies (see
Table~\ref{table:1}). We also note that fitting the SED of this galaxy
with the AGN template by Polletta~et~al.~(2006) returns the same value
of the photometric redshift as with stellar templates.

Galaxy \# 24279 has a S\'ersic index of intermediate value n$\sim$1.6,
suggesting that this is probably a bulge dominated disk galaxy. We
note that Stockton~et~al.~(2008) and McGrath~et~al.~(2008) have found
that some passive galaxies at $z>1.5$ have have disk
morphology. Finally, to obtain a reasonable fit to the light profile
of galaxy \# 24626, we had to use a combination of two components: one
with a very high S\'ersic index to reproduce the inner part of the
galaxy, and a second with n$\sim$1 to reproduce the
outskirts. Furthermore, both components have effective radius
$r_e\sim$3 kpc, implying that even if the component with lower surface
brightness were not detected (e.g. in shallower images), the effective
radius would still be found to be $\sim 3$ kpc, significantly
larger than the other galaxies. Cases like this galaxy, therefore, do
not provide an explanation to the much smaller size of the $z\sim2$
ellipticals compared to those at $z\sim0$.

We have also analyzed the residual maps produced by subtracting the
best--fit models from the corresponding original images to investigate
the presence of diffuse low--surface brightness halos, as recently
suggested by some (Mancini~et~al.~2009; Hopkins~et~al.~2009b).  For
all galaxies, these maps (see Fig.~\ref{fig2}) show the presence of
residual structure in the innermost 0.5'' of the light profile, where
the contribution by the PSF is the largest. This is very likely the
result of the variations of the PSF across the field, which cannot be
taken into account by our average PSF. To verify this possibility we
have fitted the individual stars using the same average PSF used for
galaxies, and indeed we have found residuals similar in intensity and
morphology to those of the galaxies. There is, however, no evidence of
a halo. The only exception is the case of galaxy \# 24626, which
requires a two--component model to reproduce its light profile, an
inner bulge and an outer disk. Subtracting both of these two
components, however, leaves a complex residual map, revealing a
substantially more irregular light distribution than the other
galaxies.

To further explore the possible presence of a halo, we have stacked
all the individual images together, except galaxy \# 24626. In the
H--band the stacked image has an equivalent exposure time
$T_{exp}=78,500$ sec and reaches $1\sigma$ surface brightness
sensitivity $\mu_{H,stack}\sim$27 mag arcsec$^{-2}$, and we have
performed the GALFIT analysis on it following the same procedure used
for individual galaxies.  The stacked images and the GALFIT residuals
are shown in Figures~\ref{fig1} and~\ref{fig2}, respectively. Close
companions were not masked before stacking and were not included in
the GALFIT model. The stacked image, like the individual galaxies, is
very compact. The residual images for individual galaxies and the
stack show no evidence of a large-scale diffuse halo with less than 2\%
of the residual light falling within 2" of the source.

\begin{figure}
\begin{center}
\includegraphics[scale=.5]{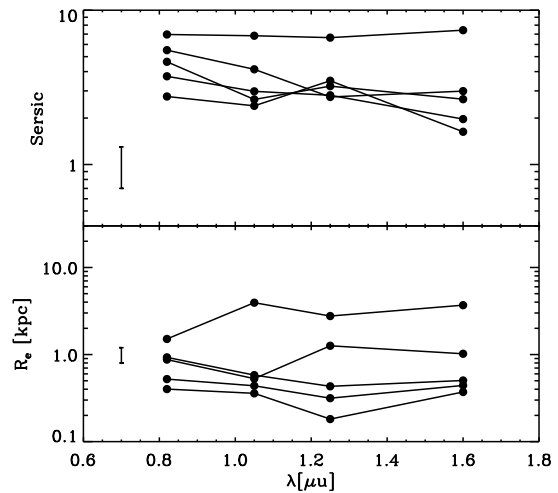}
\caption{The S\'ersic index ({\it upper panel}) and the effective radius ({\it
    bottom panel}) as a function of the wavelength. Object \# 23495, the one 
    best-fitted by a point like source, is not shown in this figure. 
    The typical error bar is marked with a vertical bar, and amounts to 
    $\sim30$\% of the measured S\'ersic indices and $\sim20$\% of $r_e$.}
\label{fig3}
\end{center}
\end{figure}

We plot in Figure~\ref{fig3} the S\'ersic index (top) and effective
radius (bottom) of the galaxies measured from the the z, Y, J and
H--band images as a function of wavelength. Assuming a mean redshift
$<z>=1.7$ for the sample, the covered rest--frame wavelength range
extends from the rest--frame UV at $\lambda\sim3300$ \AA\ to the
optical at $\lambda\sim6000$ \AA. Consistent with a visual
inspection, the two plots show weak or no morphological {\cal
k}-correction.  The S\'ersic index is constant with wavelength within
the errors, while the effective radius of three galaxies has a weak,
but statistically significant dependence on it, decreasing by about
$\sim40$\% from the z band to the H one ($r_e$ is constant for the
reminder of the sample). This weak dependence of $r_e$ with wavelength
results in a negative color gradient, with the outer parts of the
galaxies being bluer than the center. This color gradient has also
been observed by Guo~et~al.~(in preparation), who report that SED
fitting implies that the stellar populations in the outer regions of
these galaxies are, on average, younger by $\approx 0.5$ Gyr than
those in the center. Finally, we note that the observed weak
dependence of morphology on wavelength is in good quantitative
agreement with other studies of early--type galaxies at high--redshift
from ACS and NICMOS images (McCarthy~et~al.~2007;
Trujillo~et~al.~2007) as well as at lower redshift
(Papovich~et~al.~2003; Cassata~et~al.~2005).

\section{The mass-size relation at $z\sim$2}
\begin{figure}
\begin{center}
\includegraphics[scale=.48]{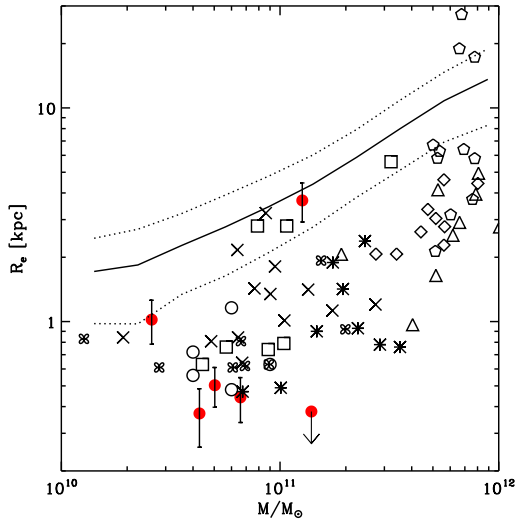}
\caption{The effective radius as a function of stellar mass for early--type
  galaxies. Our sample is marked with red filled circles, with galaxy \# 23495
  shown as an upper limit, since it is unresolved. Crosses, circles, clovers,
  squares, stars, triangles, pentagons and diamonds represent the galaxies
  from Cimatti~et~al.~(2008), Zirm~et~al.~(2007), Toft~et~al.~(2007),
  Daddi~et~al.~(2005), Van Dokkum~et~al.~(2008), Trujillo~et~al.~(2006),
  Mancini~et~al.~(2009) and Longhetti~et~al.~(2007), respectively. The
  continuous and dashed lines represent the local relation and its $1-\sigma$
  scatter, from Shen~et~al.~(2003). }\label{fig4}
\end{center}
\end{figure}

In Figure~\ref{fig4} we compare the relationship between stellar mass and size
in the H band for the six galaxies in our sample to other measures from the
literature, both at high redshift and in the local universe. When needed, we
have homogenized mass measures to M05 and CB07 with Salpeter IMF by applying
offsets\footnote{$\log(M_{CB07,M05})=\log(M_{BC03})-0.2$ and $\log(M_{Salp})=
  \log(M_{Chab})+0.25$} computed by
Salimbeni~et~al.~(2009). The figure shows that four out of our six galaxies
are located below the local relation for early--type galaxies. The remaining
two galaxies (\#19389  and \#24626) are on the local relation.

Thus, on the one hand our small sample of $\sim2$ massive and passive
galaxies seems to confirm that a significant fraction of these systems are
$\sim3$ to 5 times smaller, and hence have $\sim50$--100 times denser
central stellar density, than their local counterparts, as previously reported
from similar but shallower and/or lower quality images (Trujillo~et~al.~2007;
Toft~et~al.~2007; Zirm~et~al.~2007; Longhetti~et~al.~2007;
Van~Dokkum~et~al.~2008; Toft~et~al.~2009) or at UV rest-frame wavelengths
(Daddi~et~al.~2005; Cimatti~et~al.~2008). On the other hand, it also suggests
a diversity of morphological properties (size and stellar mass density) among
these galaxies, with some of them being similar to the local ones, which still
remains to be characterized from a statistical point of view. We plan to
report soon on one such study.

\section{Summary and Conclusions}

Massive galaxies ($M\sim10^{11}$ M$_{\odot}$) with early spectral type
at $z\sim2$ appear to often have a relatively smooth and regular
morphology, nearly independent of the wavelength, and with light
profile which is well described by a S\'ersic model, i.e. their
morphology can be classified as ``elliptical'', although systems with
an exponential component are also observed. This suggests that a tight
correlation between spectral and morphological properties similar to
what is observed in the local universe, i.e. the back bone of the
Hubble sequence, is already in place among bright galaxies at
$\sim25$\% of the cosmic time. The S\'ersic analysis has also
quantitatively shown that the morphology of these galaxies depends
very mildly on the wavelength, between the rest--frame UV and optical
windows.

One of our six galaxies is almost certainly an AGN, while another with
a spatially unresolved nuclear component is a candidate one. Such kind
of contamination should not be surprising given the similarity between
the SED of obscured AGN and quiescent galaxies. By the same token,
however, it is also possible that both types of sources contribute to
the luminous output of the same galaxy, a fact that we have not
directly investigated in this work. The apparently high level of
contamination (33\% of our sample) observed at $z\sim2$, the cosmic
era when both AGN and star--formation activity both reach their
maximum (Ueda~et~al.~2003), raises the intriguing question of whether
this is direct evidence of a link between the AGN and the mechanisms
responsible for the cessation of star formation in the galaxies. We
plan to report on this issue using larger samples from imminent new
observations with WFC3 of the GOODS fields. The high AGN contamination
is also likely to bias statistical conclusions made from
UV/Optically--selected samples of passive galaxies at $z\sim2$,
unless sensitive X--ray and mid--IR photometry (Daddi~et~al. 2005,
2007), or at least multi--wavelength high--resolution {\it HST}
imaging, can be used (like in the GOODS) to cull these sources.  In
surveys where such information is not available, like some recent
ground--based ones (e.g. van Dokkum et al. 2009), the AGN so
frequently encountered among the galaxies cannot be recognized.

Four out of six of our $z\sim2$ ellipticals lie below the local
mass--size relation for early--type galaxies, in agreement with
previous findings that they can be up to $\sim3$--5 times more
compact at given mass, and thus $\sim30$--100 times denser, than
their local counterparts.  Of the remaining two galaxies, however, one
(\# 19389), which has an unresolved central component, is at the
boundary of the local relation, while the other (\# 24626) is well
within it. We have not found any evidence of a low--surface brightness
halo surrounding these galaxies, either in the
individual images, or in their stacked one. At most, only $<2$\% of
the light (and hence of the mass, assuming a constant M/L ratio) can
be segregated in such a halo. Galaxy \# 24626, requires a
two--component light profile, one S\'ersic model with high $n$
(spheroid) embedded in a disk, to provide a good fit to the
observations, but both have the same, relatively large, effective
radius, $r_e\sim3$ kpc. Thus, what is emerging here is that these
galaxies are characterized by a dispersion of properties that also
includes systems similar to the local ones, as some have already
suggested (Cimatti~et~al.~2008; Mancini~et~al.~2009), which however
still needs to be studied and characterized. Until this work is done,
whether and/or how the ultra--compact $z\sim2$ ellipticals evolve
into the local ones or whether ultra--compact systems are also present
in the local universe but have so far eluded detection will remain
open questions.

\begin{acknowledgements}
We acknowledge the anonymous referee for the useful comments.
PC acknowledges support from NASA grant HST-GO-09822.45-A. 

\end{acknowledgements}


\begin{thebibliography}{}

\bibitem[Arnouts et al. (2007)]{arnouts06} Arnouts, S. et al. 2007, 
A\&A, 476, 137
\bibitem[Beckwith et al. (2006)]{beckwith06} Beckwith, S. V. W. et al. 2006, 
\aj, 132, 1729
\bibitem[Buitrago et al. (2008)]{buitrago08} Buitrago, F., et al. 2008, 
\apj, 687, 61L
\bibitem[Bundy et al. (2006)]{bundy06} Bundy, K., et al. 2006, 
\apj, 651, 120
\bibitem[Calzetti et al. (2000)]{calzetti00} Calzetti, D., et al. 2000, 
\apj, 533, 682
\bibitem[Cassata et al. (2005)]{cassata05} Cassata, P., et al. 2005, 
A\&A, 357, 903
\bibitem[Cimatti et al. (2008)]{cimatti08} Cimatti, A., et al. 2008, 
A\&A, 482, 21
\bibitem[Daddi et al. (2004)]{daddi04} Daddi, E., et al. 2004, 
\apj, 617, 746 
\bibitem[Daddi et al. (2005)]{daddi05} Daddi, E., et al. 2005, 
\apj, 626, 680 
\bibitem[Daddi et al. (2007)]{daddi07} Daddi, E., et al. 2007, 
\apj, 670, 163 
\bibitem[Fontana et al. (2006)]{fontana06} Fontana, A.,
et al. 2006, A\&A, 459, 745
\bibitem[Fontana et al. (2009)]{fontana09} Fontana, A.,
et al. 2009, A\&A, 501, 15
\bibitem[Giavalisco et al. (2004)]{giavalisco04} Giavalisco, M.,
et al. 2004, \apj, 600, L93
\bibitem[Hopkins et al. (2009a)]{hopkins09} Hopkins, P. F., et al. 2009, 
MNRAS, 401, 1099
\bibitem[Hopkins et al. (2009b)]{hopkins09} Hopkins, P. F., et al. 2009, 
MNRAS, 398, 898
\bibitem[Ilbert et al. (2009)]{ilbert09} Ilbert, O., et al. 2009, 
\apj, 709, 644
\bibitem[Kellermann et al. (2008)]{kellermann08} Kellermann, K. I., et
al. 2008, \apjs, 179, 71
\bibitem[Kormendy et al. (2009)]{kormendy09} Kormendy, J., et al. 2009, 
\apjs, 182, 216
\bibitem[Laidler et al. (2007)]{laidler07} Laidler, V. G., et al. 2007, 
PASP, 119, 1325
\bibitem[Longhetti et al. (2007)]{longhetti98} Longhetti, M., et al. 2007, 
MNRAS, 374, 614
\bibitem[Luo et al. (2008)]{luo08} Luo, B., et al. 2008, 
\apjs, 179, 19
\bibitem[Maraston et al. (2005)]{maraston05} Maraston, C., 2005, 
MNRAS, 362, 799
\bibitem[McGrath et al. (2008)]{mcgrath08} McGrath, E. J., et al. 2008, 
\apj, 682, 303
\bibitem[Mancini et al. (2009)]{mancini09} Mancini, C., et al. 2009, 
MNRAS, 401, 933
\bibitem[Miller et al. (2008)]{miller08} Miller, N. A., et al. 2008, 
\apjs, 179, 114
\bibitem[Naab et al. (2009)]{naab09} Naab, T., et al. 2009, 
\apj, 690, 1452
\bibitem[Papovich et al. (2003)]{papovich} Papovich, C., et al. 2003, 
\apj, 598, 827
\bibitem[Peng et al. (2002)]{Peng} Peng, C. Y., et al. 2002, 
\aj, 124, 266
\bibitem[Polletta et al. (2006)]{Polletta} Polletta, M., et al. 2006, 
\apj, 642, 673
\bibitem[Ravindranath et al. (2006)]{Ravindranath} Ravindranath, S., et al. 2006, 
\apj, 652, 963
\bibitem[Renzini (2006)]{Renzini2006} Renzini, A., 2006, 
ARA\&A, 44, 141
\bibitem[Salimbeni et al. (2009)]{Salimbeni2009} Salimbeni, S., et al. 2009, 
American Institute of Physics Conference Series, Vol. 1111, 207
\bibitem[Scarlata et al. (2007)]{Scarlata2007} Scarlata, C., et al. 2007, 
\apjs, 172, 494
\bibitem[Shen et al. (2003)]{Scarlata2003} Shen, S., et al. 2003, 
MNRAS, 343, 978
\bibitem[Stockton et al. (2008)]{Stockton08} Stockton, A., et al. 2008, 
\apj, 672, 146
\bibitem[Stoughton et al. (2002)]{Stoughton2002} Stoughton, C., et al. 2002, 
\aj, 123, 485
\bibitem[Toft et al. (2007)]{Toft2007} Toft, S., et al. 2007, 
\apj, 671, 285
\bibitem[Toft et al. (2009)]{Toft2009} Toft, S., et al. 2009, 
\apj, 705, 255
\bibitem[Trujillo et al. (2006)]{Trujillo06} Trujillo, I., et al. 2006, 
MNRAS, 373, L36
\bibitem[Trujillo et al. (2007)]{Trujillo07} Trujillo, I., et al. 2007, 
MNRAS, 382, 109
\bibitem[Ueda et al. (2003)]{ueda03} Ueda, Y., et al. 2003, 
\apj, 598, 886
\bibitem[Van Dokkum et al. (2008)]{VanDokkum08} Van Dokkum, P. G., et
al. 2008, \apjl, 677, L5
\bibitem[Van Dokkum et al. (2009)]{VanDokkum09} Van Dokkum, P. G., et al. 2009,
PASP, 121, 2
\bibitem[Valentinuzzi et al. (2009)]{valentinuzzi09} Valentinuzzi, T., et al. 
2009, [astro-ph/0907-2392]
\bibitem[Vanzella et al. (2008)]{vanzella08} Vanzella, E., et al. 
2008, A\&A, 478, 83
\bibitem[Zirm et al. (2007)]{Zirm} Zirm, A. W., et al. 2007, 
\apj, 656, 66
\end{thebibliography}
\end{document}